# A Semi-quantitative Covid-19 Individual Risk Model

Jens Braband, TU Braunschweig

Hendrik Schäbe, TÜV Rheinland

## Abstract

This paper introduces a new basic risk model that could also be utilized by Covid-19 warning apps a priori, before an action is performed. Today the common warning apps estimate risk a posteriori and give no advice on particular scenarios. The new model also has the advantage that the individual risks behind the decision-making process would be uniform (in contrast to some current regulations) and it could help to understand the risks better and could also help to reduce risks a priori. It could be easily implemented on a single app screen, needing only some individual preferences to be set and a handful of adjustments to the particular scenario that shall be assessed.

The disadvantage as of any simplified semi-quantitative risk models is that calibration is not easy (as some calibration points may even contradict) and that cumulative effects are hard to integrate e. g. the joint effect of combined scenarios. But, in principle calibration is feasible and it may be a good decision to calibrate the model conservatively.

## Introduction

In Germany, an app has been introduced to cope with the Covid-19 epidemic. This app computes a risk for persons to have been infected caused by contacts. We have already analyzed that this risk model is incomplete [1,2]. But the major flaw is that the app computes risk a posteriori. This is similar in most Covid apps, e. g. also the UK version, which has sent hundred thousands into quarantine recently, uses a similar model [9]

What is needed instead is a proactive risk estimate that helps to prioritize actions of individuals in everyday situations in order to reduce the number of dangerous encounters a priori. There exist already simulators to calculate the risk for particular situations e. g. indoor meetings [7], but some of the models are not easy to use as many parameters have to be estimated or the result is not intuitive, e. g. the probability of a Covid-19 infection as a result of the scenario. But what does a say 3.7% probability of a Covid-19 infection mean. Should the risk be taken, or not? And finally, these models do not take into account individual parameters, as the same scenario may be acceptable for a student, but not for an 80-year old person.

The approach to construct such semi-quantitative risk models is well known [5] and the concept has been successfully applied in the railway domain [3]. The major challenges are the composition of the model, the scaling of the parameters and the calibration of the risk matrix.

# The Basic Risk Model

The general assumption is that the main and dominating way of infection is by spreading of aerosols contaminated with Covid-19 [7]. Other ways of infection are neglected in the model e.g. by contact to contaminated surfaces. However, it seems easier to cover these ways by simple rules, e. g. disinfection.

We construct an individual risk model stimulated by similar models from dosimetry. As all estimates are only approximate, we define a semi-quantitative risk model similar to the well-known Risk Score Matrix (RSM) [3], that is composed of three main parameters: Severity, Frequency and risk reducing Barriers.

The first observation is that the impact on an individual is well represented by the severity S of the clinical course of the infection. On average this is represented by the risk groups defined by RKI [4], ranging from I to VI. These are defined as the priority groups for vaccination. Note that if a person is exposed regularly to more vulnerable persons, but this case is not covered by the classification, then this more restrictive impact category should be chosen.

| Severity S | Description |
|---|---|
| I | Persons older than 80, inhabitants of nursing homes, medical personal or care personal with very high exposure or contact to vulnerable persons… |
| II | Persons older than 75, persons with dementia or mental handicap, medical personal or care personal with high exposure or contact to vulnerable persons… |
| III | Persons older than 70, persons with severe medical indications, medical personal or care personal with moderate exposure or contact to vulnerable persons… |
| IV | Persons older than 65, persons with moderate medical indications, medical personal or care personal with low exposure or contact to vulnerable persons, teachers… |
| V | Persons older than 60, persons in system relevant jobs, … |
| VI | All other groups |

Table 1: Risk groups based on RKI recommendations [4]

Secondly the frequency F (or probability) of infection is proportional to the mean cumulative dose of virus material to which the individual is exposed. Here we exploit the similarity that we have sources of activity (the number N of persons the individual is exposed to), which transmit virus material with a certain intensity I (which is assumed to be constant at least temporarily e. g. over a day) and the duration of the exposure T. We need to take into account the cumulative effect C of some activities that are carried out regularly or repeatedly. Note that the number N may only include persons that have an unknown or outdated test or vaccination status. Persons with a current negative test or complete vaccination need not to be included. Also, persons of the own household do not need to be included, so N represents the number of untested persons external to the household that are met.

Finally, the exposed persons can reduce their risk by introducing risk reducing barriers that are assumed to be the fixed (or average) distance D, the type M of masks worn and the ventilation V (relating to the location where the activity takes place).

The intensity I could be defined e. g. by virus material per volume or surface e. g. [M/m$^2$], D would be measured by [m] and T by [min]. For the risk model we would have to take some assumptions how the intensity decreases with distance to the source.

Several assumptions on the dependence of concentration depending on distance can be made:

a) Thinning of the virus in a volume. The concentration of the virus is the amount of the virus per volume unit, assuming more or less equal distribution. Then the concentration would decrease with D$^3$, where D is the distance between the source (infected person) and the receiver. This assumption would only hold, if there is sufficient convection
b) The virus is emitted as a spherical wave, e. g. because of playing a brass instrument, sneezing etc.  Then the virus is approximately present on the surface of a sphere. So, the concentration would decrease with D$^2$.
c) The virus is emitted as a cylindrical wave. Then the concentration would decrease proportional to D. This can be the case if e. g. infected singers are standing on a stage, emitting their virus into a large room
d) The virus is spread in the room, reflected from the walls, so that an equal concentration occurs in the entire room. Then, distancing would not decrease the virus concentration.

Therefore, we think that assuming a dependence of the intensity I on 1/D is a plausible assumption, which should cover typical cases or be conservative.

We may also define a nominal virus intensity I$_0$ at a given distance, say 1 m. This would still depend on the infectiousness of the person, or on average the weekly incidence W (per 100,000 inhabitants over a week). So, in place of using the infectiousness of the single persons I and the dependence on distance etc, we will only use the parameter W as an average value

For the model we may assume that F is proportional to the product of N times W times C times T. To arrive at a semi-quantitative representation, we can take logarithms corresponding to an adequate basis, say $\sqrt{10}$, so that on this scale (and adequate transformation) [5]

$$F = N + W + C + T.$$

For each parameter, a table is calibrated approximately according to this scale.

Note that at large also the fatality rates between some of the risk groups follows this scaling, e. g. between the over-80s, over-70s, over-60s and younger patients [6]. Also, for Germany there is some evidence to support this scaling approximately [8].

| Number N | Description |
|---|---|
| 1 | Single Person |
| 2 | Several people e. g. a couple or small group |
| 3 | A large group e. g. 10 persons |

| 4 | Many people e. g. 30 persons |
| 5 | Very many people e. g. 100 persons |

Table 2: Number of people met

| Weekly Incidence W | Description |
|---|---|
| 1 | Very low e. g. below 10 |
| 2 | Low e. g. below 35 |
| 3 | Moderate, e. g. below 100 |
| 4 | High, e. g. below 300 |
| 5 | Very high |

Table 3: Weekly incidence

| Cumulative effect C | Description |
|---|---|
| 0 | Exposure once per week or less |
| 1 | Exposure several times per week |
| 2 | Daily exposure or more often |

Table 4: Cumulative exposure

| Exposure Time T | Description |
|---|---|
| 1 | Very short, below 1 min |
| 2 | Short, e. g. below 5 min |
| 3 | Medium e. g. below 10 min |
| 4 | Long, e. g. below 30 min |
| 5 | Very long, e. g. below 90 min |
| 6 | Above |

Table 5: Single exposure

It seems reasonable to set a threshold for N, so if a higher number of persons is present, then this activity should not be performed. The main reason is that such activities may be super-spreader vents that may have a different, maybe exponential dynamic.

The risk is now presented by the set of the parameters S (severity) and F. These parameters can be presented in a risk score matrix. However, the matrix still needs to be calibrated, i.e., all combinations of the parameters S and F need to be classified into the following color coded risk classes:

      Green – the risk is acceptable,

      Yellow – the risk is undesirable and should be avoided, if alternatives exist,

      Orange – the risk is undesirable and should be avoided,

      Red – the risk is not acceptable.

## Calibration of the Risk Model

As a first step we may calibrate a risk score matrix, that represents the parameters S and F. Calibration may be performed by assessing activities that are allowed for the different weekly incidences W.

As a calibration point, we propose a risk that is neglected by the risk model of the German Corona Warning App: if the duration of a single encounter with an infected person is below 10 minutes, then it is neglected and only in later versions of the app the cumulative effect of such encounters was regarded. But we may take T=3, C=1 and N=1 as possible calibration parameters. The important parameter is W, which is not regarded in the Corona warning app, but we may assume that the risk is acceptable for W=1 or even W=2. So conservatively, we choose F=6. And this risk should be acceptable for any severity, also S=I.

An alternative calibration point is that even for high incidences, say W=3, it is currently accepted that one household meets a single person (N=1) as often as they like (C=2) for an undefined time (but on average we assume T=5), so that F=11. This may also be indoor, without ventilation or any masks (D=A=V=0). But we may safely assume that this holds mainly for persons in the lowest severity category S=VI.

Also, scenarios that are forbidden may serve as calibration points. E. g. it was strictly forbidden to visit any nursing home or hospitals, so even S=IV (for hospitals), N=1, C=0, T=3 was forbidden even with M=2, D=1 and V=1, which would have been possible to arrange in most cases. However, it is easy to see, that such a calibration point conflicts with the other points. So probably this interdiction was caused by other organizational reasons (e. g. safeguarding the hospital not the individuals) or chain effects that this simple model can't cover.

Now a table may be filled either by other calibration points, by interpolation or applying generally accepted risk analysis principles. Here, we use the calibration points specified above and elementary logic, i.e., that increasing the frequency or the severity must increase the risk and that jumps of the risk must be avoided.

We must also note that the risk model is somewhat simplistic:It does not take into account the infectiousness of the surrounding persons, since only the weekly incidence W is used. Therefore, the model cannot be applied in situations, where people have contact to infected persons with larger probability, e.g., on Covid stations in hospitals – the risk would largely be underestimated or even in situations, where people normally do not have contact with infected persons, even with high W values, which might be the case with very cautious persons, who do a personal screening of their contacts.

All parameters have been discretized as is done normally in semi-quantitative models.

The severity S includes some consideration on the exposure to infected persons, which is against the logic of a risk score model. However, we have used these severity classes to be consistent with the classes used by the RKI.

Some protection measures (barriers) are not yet taken into account in the model.

| F \ | I | II | III | IV | V | VI |
|---|---|---|---|---|---|---|
| 3 | | | | | | |
| 4 | | | | | | |
| 5 | | | | | | |
| 6 | 🟢 | | | | | |
| 7 | 🟡 | 🟢 | 🟢 | | | |
| 8 | 🟠 | 🟡 | 🟡 | | | |
| 9 | 🔴 | 🟠 | 🟠 | 🟢 | 🟢 | |
| 10 | 🔴 | 🔴 | 🔴 | 🟡 | 🟡 | |
| 11 | 🔴 | 🔴 | 🔴 | 🟠 | 🟠 | 🟢 |
| 12 | 🔴 | 🔴 | 🔴 | 🔴 | 🔴 | 🟡 |
| 13 | 🔴 | 🔴 | 🔴 | 🔴 | 🔴 | 🟠 |
| 14 | 🔴 | 🔴 | 🔴 | 🔴 | 🔴 | 🔴 |
| 15 | 🔴 | 🔴 | 🔴 | 🔴 | 🔴 | 🔴 |

Table 6: Proposed Risk Score Matrix

Table 6 represents the risk score matrix. The final recommendation from this matrix can be expressed by a simple color code:

- Green = risk may be accepted if all other recommended measures are carefully implemented
- Yellow = risk should be avoided if alternatives exist
- Orange = risk should be taken only if it is unavoidable e. g. as part of necessary work
- Red = risk should be taken only in exceptional circumstances e. g. emergency

## Introducing barriers

Finally, barriers D, M and V can be assessed to reduce the risk. These parameters may be mostly voluntarily chosen or influenced by the individual, e. g. the time of the meeting D may be reduced, or the location may be changed, or masks can be chosen. These parameters directly reduce the risk by the scores represented in the tables. The magnitude of the scores can be validated by more complex simulation models [7]. Note that also other parameters from such models could be considered, e. g. the volume of the space where the persons meet, or the type of activity, but this makes the model more complex or the parameters harder to estimate.

| Distance D | Description |
|---|---|
| 0 | Very low e. g. below 2m |
| 1 | Low e. g. below 5m |
| 2 | Large |

Table 7: Average distance

| Mask M | Description |
|---|---|
| 0 | None |
| 1 | Simple everyday mask |
| 2 | Medical mask e. g. FFP2 |
| 3 | Better |

Table 8: Type of masks

| Ventilation V | Description |
|---|---|
| 0 | None |
| 1 | Frequent ventilation e. g. open windows |
| 2 | Air condition with filters |
| 3 | Outdoor |

Table 9: Ventilation conditions

Taking into account the barriers, we arrive at the following modified formula for F:

$$F = N + W + C + T - D - M - V$$

The risk score matrix given in table 6 is still valid.

# Examples

Assume that a 55-year-old without any medical preconditions or contact to vulnerable contact persons (S=VI) wants to go for click&meet shopping in 3 shops (C=1). There may be many people in the shop including shopkeepers (N=4), exposure is long (T=2). The weekly incidence in the town, where the shops are located, is moderate (W=3). So, the score F=10 would be acceptable without any additional measures. Still the advice to wear masks should be followed.

Now assume that our test person would be a nurse in direct contact to vulnerable persons, (S=I), but wit the same conditions as in the paragraph above. Here F=10 would be unacceptable, but we could reduce the risk by wearing FFP2 mask (M=2) and either reducing the shopping tour (C=0) or shortening the time spent in the shops (T=1) or smaller shops (N=3). This would reduce F to 8.

Assume additionally, that our nurse takes the metro to work (C=2). If the train is crowded (N=4), the trip is long (T=3) and a FF2 mask is worn (M=2), then this behavior would not be acceptable even for low incidence (say W=2), since we would have F=9.

Taking a look at schools, assume a school is implementing the so-called AB model (C=1, N=3) with normal hours (T=6), where classes are split into two subgroups that are separated. Then we need risk reduction e. g. open windows (V=1) and masks (M=1) even for low incidences (up to W=3) and normal health (S=VI). The situation might be different for students with medical conditions or teachers (S=4). So, teachers might be advised to wear better masks or keep a larger distance, or they should be prioritized for vaccination

From the risk score matrix also another fact becomes evident. If two different groups with different severity classes are in the same situation, i.e., with the same F, the group with the smaller S value has a larger risk. Then, risk reduction measures need to be taken by all persons that are in this situation. Also, it becomes clear that vaccination persons with a smaller S value would immediately give a larger risk reduction than if persons in a category with a larger S value are vaccinated. The same considerations also motivate measures to vaccinate persons with priority that reach principally a larger F value, caused by their (meeting many persons) working or (crowded) living situation.

# Summary

This paper introduces a new basic risk model that could also be utilized by Covid warning apps a priori, before an action is performed. It has the advantage that the individual risks behind the decision-making process would be uniform (in contrast to some current regulations) and it could help to understand the risks better and could also help to reduce risks. It could be easily implemented on a single app screen, needing only some individual preferences to be set (e. g. the severity category) and a handful of adjustments to the particular situation.

The disadvantage as of any simplified semi-quantitative risk models is that calibration is not easy (as some calibration points may contradict) and that cumulative effects are hard to integrate e. g. the joint effect of shopping and metro rides. But, in principle the calibration is feasible, and it may be a good decision to calibrate the model conservatively. Also, the model covers only infections by spreading of aerosols, but this seems in line with current research.

It is also generally known that in many situations simple models are more robust than more sophisticated models [10], because more sophisticated models are often more trimmed to particular scenarios or data, e. g. due to overfitting, and that the generalization quality of simple, e. g. linear models, is often similar to more complex, e. g. nonlinear, models.